\documentclass[useAMS,usenatbib]{mn2e}

\title[A Polarimetric Method for Measuring Black Hole Masses in Active Galactic Nuclei]{A Polarimetric Method for Measuring Black Hole Masses in Active Galactic Nuclei}
\author[M.Yu. Piotrovich, Yu.N. Gnedin, N.A. Silant'ev,  T.M. Natsvlishvili, S.D. Buliga]{M.Yu. Piotrovich\thanks{E-mail: mike@gao.spb.ru}, Yu.N. Gnedin\thanks{E-mail: gnedin@gao.spb.ru}, N.A. Silant'ev, T.M. Natsvlishvili, S.D. Buliga\\
Central Astronomical Observatory at Pulkovo, Saint-Petersburg, Russia.}

\begin{document}

\date{Accepted for publication in MNRAS}

\pagerange{\pageref{firstpage}--\pageref{lastpage}} \pubyear{2015}

\maketitle

\label{firstpage}

\begin{abstract}
The structure of the broad emission line region (BLR) in active galactic nuclei (AGN) remains unclear. We test in this paper a flattened configuration model for BLR. The virial theorem, by taking into account the disc shape of BLR, allows us to get a direct connection between the mass of a supermassive black hole (SMBH) and the inclination angle of the accretion flow. The inclination angle itself is derived from the spectropolarimetric data on broad emission lines using the theory for the generation of polarized radiation developed by Sobolev and Chandrasekhar. As the result, the new estimates of SMBH masses in AGN with measured polarization of BLR are presented. It is crucial that the polarimetric data allow also to determine the value of the virial coefficient that is essential for determining SMBH masses.
\end{abstract}


\section{Introduction}

It is generally accepted that AGNs are powered by accretion to a SMBH and the broad emission lines seen in type I AGN are produced as the result of photoionization of gas in the BLR. However the structure and kinematics of the BLR is a key problem in modern astrophysics. Broad emission lines are emitted in the vicinity of SMBH in AGN, but this region is not resolved in observations. Properties of the broad emission lines are used to estimate the mass of the central SMBH. The traditional method for estimating the SMBH mass is to use the virial theorem. It means that the mass of the SMBH can be estimated as \citep{vestergaard06,fine08}:

\begin{equation}
 M_{BH} = f \frac{R_{BLR} V_{BLR}^2}{G},
 \label{eq01}
\end{equation}

\noindent where $M_{BH}$ is the mass of a black hole, $f$ is a virial parameter that defines the geometry, velocity field and orientation of BLR, $R_{BLR}$ is the radius of BLR and $V_{BLR}$ is the velocity dispersion that is measured as the full width of the emission line in the variable spectrum. The BLR radius $R_{BLR}$ is usually given by the time delay between continuum and emission line variations.

There are various approaches for determining the value of $f$. \citet{labita06} found that in quasars an isotropic BLR fails to reproduce the observed line widths and shapes, and a disk model is preferred. A disk like geometry for the BLR has been proposed by several authors \citep{decarli08}. Some authors suggested that the BLR can not be completely flat (e.g. \citet{collin06}). It means that disk may have a finite half thickness $H$, or a profile with $H$ increasing more than linearly with the disk radius. Other models propose the existence of warped disks \citep{tremaine14}.

According to \citet{kollatschny06,kollatschny13} the hydrogen lines are emitted in a more flattened configuration in comparison to the highly ionized lines. $H_\beta$ lines originate at heights of 0.7 to 1.6 light days and at distances of 1.4 to 2.4 light days with height/distance ratios $H/R \sim 0.07 - 0.5$. \citet{pancoast13} found that the geometry of the BLR is generally a thick disk viewed close to face on. \citet{eracleous94} have found that the inclination angle of BLR is $24^{\circ} - 30^{\circ}$, and \citet{eracleous96} suggested that the inclination of the BLR can be $i = 19^{\circ} - 42^{\circ}$.

The virial coefficient $f$ depends strongly on the BLR geometry, velocity field and orientation. Many authors used the value $f \sim 1$. \citet{peterson99} found $f = 3/4$. \citet{onken04} found the mean value of the virial coefficient $f = 1.4$. \citet{mclure01} have shown that for a disk inclined at an angle $i$ to the observer the virial coefficient value is

\begin{equation}
 f = \frac{1}{4 \sin^2{i}}.
 \label{eq02}
\end{equation}

In this paper we adopt the disk like model for the BLR of Seyfert galaxies and the expression for $f$ given by (\ref{eq02}). The value of the inclination angle can be determined from polarimetric observations using the standard Chandrasekhar-Sobolev theory \citep{chandra50,sobolev63} of multiple scatterings of the radiation on free electrons and Rayleigh scattering on gas molecules and small dust particles. According to these classical works, the polarization degree of scattered radiation depends strongly on the inclination angle. The scattered radiation has the maximum linear polarization $P_l = 11.7\%$ when the line of sight is perpendicular to the normal to the semi-infinite atmosphere (Milne problem). \citet{chandra50} and \citet{sobolev63} presented the solution of the so-called Milne problem, i.e. multiple scattering of light in optically thick flattened atmospheres. The Milne problem corresponds to the propagation and scattering of light in optically thick disk-like region, i.e. this solution can be applied to BLR. The idea of inferring the inclination of a black hole accretion disk from observation of its polarized continuum has been suggested by \citet{li09} and developed for determining the orientation of the X-ray producing inner region of the accretion disk around a black hole in X-ray binary systems. Another idea for using spectropolarimetric data has been suggested by \citet{afanasiev11}. The authors used the spectropolarimetric observation of the line profile for Sy 1.5 galaxy Mrk~6 and assumed that the scattering region is located in the inner part of the torus. Recently \citet{marin14}, based on archival data, reported the first compilation of 53 estimated AGN inclinations matched with ultraviolet/optical continuum polarization measurements.

But these data are obtained in the widebands of wavelength and do not include the data on the broad $H_{\alpha}$ and $H_{\beta}$ emission that are used for determination of the masses of SMBH. The values of the inclination angles are obtained in many cases by indirect methods. For example, \citet{wu01} and \citet{zhang02} estimated the orientation angles $i$ assuming a definite mass/velocity dispersion between AGN and regular galaxies where AGN are located. For some Seyfert galaxies the inclination estimations are obtained from optical polarization data that are produced in continuum and can not be directly related to BLR. According to \citet{tremaine14} the warped disks can exist in AGN and it means that BLR and the standard accretion disk, located closely to SMBH, can have different inclinations. This phenomenon is the reason why we used the polarimetric data of $H_{\alpha}$ emission line presented by \citet{smith02}.

We use the theory of multiple scatterings of polarized radiation \citep{chandra50,sobolev63} and the disk like model for the BLR. As the result we estimate the values of the virial factor and the mass values for SMBH in AGN. {\underline We have in our order the detailed atlas of the values of the polarization degree and its dependence on the inclination angle $i$ with the step of 0.005 for $\mu = \cos{i}$}. Besides it is convenient in some cases to use the analytical formula for the polarization degree $P_l(\mu)$ obtained in \citet{silantev10}:

\begin{equation}
 P_l(\mu) = 11.7\% \frac{1 - 2.2399\mu + 2.9377\mu^2 - 1.6978\mu^3}{1 + 2.2001\mu + 0.062\mu^2 - 0.1988\mu^3}.
 \label{eq03}
\end{equation}

The determination of the SMBH mass depends strongly on the virial coefficient. According to Eq.(\ref{eq02}) the virial coefficient depends on the inclination angle. The calculations of \citet{chandra50} and \citet{sobolev63} of the degree of polarization of multiple scattered radiation coming out from the flattened atmosphere is strongly dependent on the inclination angle $i$. Thus, according to Eq.(\ref{eq02}) the determination from polarization data the value of the inclination angle allows us to determine the virial coefficient $f$.

\section{Determining the virial coefficient}

For determining the virial coefficient we use Eq.(\ref{eq02}). The polarimetric data that are necessary for determining the virial coefficient are presented in the spectropolarimetric atlas of \citet{smith02}. They obtained the values of polarization degree and position angles for 36 Type I Seyfert galaxies during a number of different runs at the Anglo-Australian and William Herschel telescopes. From 36 objects presented in the atlas of \citet{smith02} approximately 13 AGNs have the equal polarization degree for $H_\alpha$ line and continuum and approximately equal values of the position angle within the error limits. For most of the observed objects there is difference between the values of the polarization degree and the position angle. This fact can testify to the difference in the inclinations between disk shaped BLR and the accretion disk that can be described by Shakura-Sunyaev model \citep{shakura73}. We shall show below that this conclusion is confirmed by polarimetric observations. Continuum polarization degree of AGN from Palomar-Green catalogue have been measured on 6-m telescope of Special Astrophysical Observatory \citep{afanasiev11}.

\renewcommand{\arraystretch}{1.2}
\begin{table}
\centering
\caption{The values of the virial coefficient $f$ determined via measured polarization of Broad $H_\alpha$ emission of AGN. Polarization data from \citet{smith02}. $i$ is the inclination angle.}
\begin{tabular}{lll}
\hline
Object          & $i$ [deg]            & $f$ \\
\hline
Akn~120         & $29\pm 1$            & $1.064^{+0.036}_{-0.034}$ \\
Akn~564         & $26^{+3}_{-1}$       & $1.25^{+0.16}_{-0.17}$ \\
ESO~012-021     & $34\pm 4$            & $0.8^{+0.2}_{-0.14}$ \\
Fairall~51      & $77.5\pm 0.5$        & $0.262\pm 0.001$ \\
I Zw 1          & $26\pm 1$            & $1.315^{+0.095}_{-0.085}$ \\
ESO~141-G35     & $43.5^{+0.5}_{-1.0}$ & $0.53^{+0.032}_{-0.015}$ \\
KUV~18217+6419  & $16^{+2}_{-1}$       & $3.19^{+0.45}_{-0.59}$ \\
Mrk~6           & $41^{+0.5}_{-1.0}$   & $0.58^{+0.02}_{-0.01}$ \\
Mrk~279         & $21^{+2}_{-1}$       & $2.0^{+0.8}_{-0.4}$ \\
Mrk~290         & $29\pm 2$            & $1.07^{+0.12}_{-0.11}$ \\
Mrk~304         & $37.5^{+2.0}_{-1.5}$ & $0.665^{+0.051}_{-0.045}$ \\
Mrk~335         & $32.8^{+0.7}_{-0.8}$ & $0.86^{+0.04}_{-0.03}$ \\
Mrk~509         & $31.0^{+1.0}_{-1.5}$ & $0.94^{+0.09}_{-0.04}$ \\
Mrk~705         & $25.0^{+3.8}_{-4.0}$ & $1.05^{+0.33}_{-0.19}$ \\
Mrk~841         & $25^{+3}_{-2}$       & $1.38^{+0.29}_{-0.21}$ \\
Mrk~871         & $36.5^{+2.5}_{-4.0}$ & $0.7^{+0.16}_{-0.07}$ \\
Mrk~876         & $28.0^{+1.5}_{-3.0}$ & $1.17^{+0.21}_{-0.14}$ \\
Mrk~896         & $22^{+5}_{-6}$       & $1.74^{+1.38}_{-0.51}$ \\
Mrk~915         & $30.0^{+3.5}_{-1.0}$ & $1.0^{+0.02}_{-0.19}$ \\
Mrk~926         & $21.0^{+3.0}_{-3.5}$ & $1.95.0^{+0.83}_{-0.40}$ \\
Mrk~985         & $35.0^{+1.0}_{-0.6}$ & $0.75^{+0.04}_{-0.02}$ \\
MS~1849.2-78.32 & $50.5^{+2.5}_{-2.0}$ & $0.42^{+0.02}_{-0.02}$ \\
NGC~3516        & $18\pm 2.5$          & $2.5^{+1.27}_{-0.45}$ \\
NGC~3783        & $22.0^{+1.0}_{-1.5}$ & $1.79^{+0.26}_{-0.1}$ \\
NGC~4051        & $31\pm 3$            & $0.96^{+0.13}_{-0.15}$ \\
NGC~4593        & $33.0^{+2.5}_{-0.5}$ & $0.80\pm 0.06$ \\
NGC~5548        & $21.0^{+0.5}_{-1.5}$ & $1.98^{+0.23}_{-0.13}$ \\
NGC~6104        & $30.0^{+3.5}_{-4.0}$ & $1.00^{+0.32}_{-0.18}$ \\
NGC~6814        & $54\pm 1$            & $0.38^{+0.0.05}_{-0.10}$ \\
NGC~7213        & $21^{+7}_{-13}$      & $1.95\pm 0.8$ \\
NGC~7469        & $15^{+2}_{-1}$       & $3.73^{+0.69}_{-0.82}$ \\
NGC~7603        & $23.0^{+2.0}_{-1.7}$ & $1.63^{+0.28}_{-0.22}$ \\
PG~1211+143     & $15\pm 4$            & $3.62^{+3.73}_{-1.28}$ \\
UGC~3478        & $45.5\pm 1.5$        & $0.49\pm 0.3$ \\
WAS~45          & $61.5^{+2.0}_{-1.3}$ & $0.32\pm 0.01$ \\
\hline
\end{tabular}
\end{table}
\renewcommand{\arraystretch}{1.0}

Our results for the virial coefficient $f$ are presented at Table 1. In the second column of this table we present the values for the BLR inclinations obtained from polarimetric data of \citet{smith02}. The virial factor for most of the objects from the Table 1 is $f \sim 1$ within error limits, in accordance with \citet{mclure02}. For some objects, for example, Mrk~841, Mrk~896, Mrk~926, NGC~3516, NGC~3783, NGC~5548 the virial coefficient is in better agreement with the mean value of $f = 1.4$ \citep{onken04}. For a number of objects including Fairall~51, Mrk~6, MC~1849.2-78.32, NGC~6814, UGC~3478 and WAS~45 the inclination angle exceeds considerably the value $i = 30^{\circ}$, and these objects have values of the virial coefficient of $f < 1.0$. This fact can mean that for BLR the model of random orbits \citep{peterson99} is more realistic.

\section{Determining SMBH masses}

After deriving the virial coefficient, it is possible to determine the SMBH mass in these particular AGN using Eq.(\ref{eq01}). For a disk shape BLR, Eq.(\ref{eq01}) takes a form:

\begin{equation}
 \sin{i} = \frac{1}{2} \left(\frac{R_{BLR}}{R_g}\right)^{1/2} \left(\frac{FWHM}{c}\right),
 \label{eq04}
\end{equation}

\noindent where $FWHM$ is the observed full width of the emission line \citep{vestergaard06,wang09,ho08,feng14}, and $R_g = G M_{BH} / c^2$ is the gravitational radius. The quantity $R_{BLR} V^2_{BLR} / G$ is called the ''virial product''(VP). This quantity is based on the two observable quantities: BLR radius and emission line width and has units of mass. The dimensionless factor $f$ in Eq.(\ref{eq01}) depends on the structure, velocity field and inclination of BLR and it is different for each AGN. For the flattened, disk-like structure of BLR size is measured usually from reverberation mapping via the time lag between the broad line emission and continuum variabilities. $V_{BLR}$ is measured usually as full width at half maximum (FWHM) or as the line width that is usually characterized by the broad line dispersion. The full width of half maximum (FWHM) is used in many papers. The value of $\sin{i}$ can be derived from the data of the degree of polarization that is strongly dependent on the inclination angle. We have the results of the detailed numerical calculations of the degree of polarization $P_l(\mu)$ for the radiation scattered in the optically thick plane parallel atmosphere. These calculations are made in the framework of the classical Chandrasekhar-Sobolev theory \citep{chandra50,sobolev63} with the step for $\mu$ being equal to 0.005.

\renewcommand{\arraystretch}{1.2}
\begin{table}
\centering
\caption{The masses of SMBH in AGNs determined via measured polarization of broad $H_\alpha$ emission.}
\begin{tabular}{lll}
\hline
Object          & $M_{BH} / M_{\odot}$                & $M_{BH} / M_{\odot}$ \\
                & (observations)                      & (literature) \\
\hline
Akn~120$^2$     & $(3.23^{+0.44}_{-0.32})\times 10^8$ & $(4.49 \pm 0.93)\times 10^8$ \\
Akn~564$^1$     & $(1.2^{+0.94}_{-0.49})\times 10^6$  & $\sim 1.1 \times 10^6$ \\
Fairall~9$^1$   & $(1.34^{+0.76}_{-0.36})\times 10^8$ & $(2.55 \pm 0.56)\times 10^8$ \\
I Zw 1$^2$      & $10^{7.6\pm 0.17}$                  & $10^{7.441^{+0.093}_{-0.119}}$ \\
Mrk~6$^3$       & $(1.09^{+0.37}_{-0.25})\times 10^8$ & $(1.36 \pm 0.12)\times 10^8$ \\
Mrk~279$^2$     & $(8.13^{+1.24}_{-1.26})\times 10^7$ & $(15.2^{+3.25}_{-3.18})\times 10^7$ \\
Mrk~290$^3$     & $(3.94\pm 0.19)\times 10^7$         & $(2.43 \pm 0.37)\times 10^7$ \\
Mrk~304$^2$     & $10^{8.4^{+0.09}_{-0.02}}$          & $10^{8.511^{+0.093}_{-0.113}}$ \\
Mrk~335$^1$     & $(1.56^{+0.19}_{-0.15})\times 10^7$ & $(1.42 \pm 0.37)\times 10^7$ \\
Mrk~509$^3$     & $(1.35\pm 0.12)\times 10^8$         & $(1.39 \pm 0.12)\times 10^8$ \\
Mrk~705$^4$     & $10^{7.07^{+0.11}_{-0.09}}$         & $10^{6.79\pm 0.5}$ \\
Mrk~841$^2$     & $10^{8.55\pm 0.1}$                  & $10^{8.523^{+0.079}_{-0.052}}$ \\
Mrk~871$^4$     & $10^{7.04^{+0.09}_{-0.06}}$         & $10^{7.08 \pm 0.5}$ \\
Mrk~876$^2$     & $10^{8.57^{+0.19}_{-0.52}}$         & $10^{9.139^{+0.096}_{-0.122}}$ \\
Mrk~896$^5$     & $10^{7.07\pm 0.06}$                 & $10^{7.01}$ \\
Mrk~926$^6$     & $10^{8.8^{+0.19}_{-0.11}}$          & $10^{8.36\pm 0.02}$ \\
Mrk~985$^7$     & $3.18\times 10^7$                   & $5.71 \times 10^7$ \\
NGC~3516$^8$    & $10^{8.06 \pm 0.27}$                & $10^{7.88^{+0.04}_{-0.03}}$ \\
NGC~3783$^2$    & $10^{7.7\pm 0.11}$                  & $10^{7.47^{+0.07}_{-0.09}}$ \\
NGC~4051$^{10}$ & $(1.64^{+0.67}_{-0.55})\times 10^6$ & $(1.58^{+0.50}_{-0.65})\times 10^6$ \\
NGC~4593$^{10}$ & $(8.25^{+3.46}_{-3.06})\times 10^6$ & $(9.8 \pm 2.1)\times 10^6$ \\
NGC~5548$^2$    & $(7.84^{+0.53}_{-0.46})\times 10^7$ & $(7.827 \pm 0.017)\times 10^7$ \\
NGC~6104$^{11}$ & $10^{7.16^{+0.09}_{-0.08}}$         & $10^{7.39}$ \\
NGC~6814$^4$    & $10^{6.94^{+0.077}_{-0.09}}$        & $10^{7.02 \pm 0.5}$ \\
NGC~7213$^4$    & $10^{6.83^{+0.84}_{-0.33}}$         & $10^{6.88 \pm 0.5}$ \\
NGC~7469$^{10}$ & $10^{7.54^{+0.17}_{-0.22}}$         & $10^{7.19 \pm 0.13}$ \\
PG~1211+143$^2$ & $10^{8.34^{+0.29}_{-0.20}}$         & $10^{7.961^{+0.082}_{-0.101}}$ \\
\hline
\multicolumn{3}{l}{(1) \rule{0pt}{11pt}\footnotesize \citet{reynolds13};
                   (2) \rule{0pt}{11pt}\footnotesize \citet{vestergaard06};}\\
\multicolumn{3}{l}{(3) \rule{0pt}{11pt}\footnotesize \citet{feng14};
                   (4) \rule{0pt}{11pt}\footnotesize \citet{ho08};}\\
\multicolumn{3}{l}{(5) \rule{0pt}{11pt}\footnotesize \citet{shankar12};
                   (6) \rule{0pt}{11pt}\footnotesize \citet{winter10};}\\
\multicolumn{3}{l}{(7) \rule{0pt}{11pt}\footnotesize \citet{ziolkowski08};
                   (8) \rule{0pt}{11pt}\footnotesize \citet{wu01};}\\
\multicolumn{3}{l}{(9) \rule{0pt}{11pt}\footnotesize \citet{brenneman13};
                   (10) \rule{0pt}{11pt}\footnotesize \citet{wang09};}\\
\multicolumn{3}{l}{(11) \rule{0pt}{11pt}\footnotesize \citet{woo14};}\\

\end{tabular}
\end{table}
\renewcommand{\arraystretch}{1.0}

The results of our calculations of the SMBH masses are presented in Table 2. In the last column of the table the published values for the SMBH mass are presented. Our values show a good overall agreement with the previous determinations  of black hole masses, obtained by various methods. For a number of objects there is a difference. For Fairall~9 our estimate of the SMBH mass looks lower, but our upper limit value coincides with the low-order limit value of BH, presented by \citet{reynolds13}. For Ark~120 the situation looks as the same and there is the coincidence of our estimated value with the low-order value of BH mass, presented by \citet{vestergaard06}. The similar situation occurs also for NGC~3516, NGC~3783, NGC~5548 and Mrk~279. Obly for Mrk~290 our low-order estimate 1.34 times higher than the upper limit of BH mass, presented by \citet{feng14}.

The difference between our and previous estimates of mass for some black holes can be associated with the real determination of the virial parameter $f$ from Eq.(\ref{eq01}).  \citet{mclure01} value of $f$ (Eq.(\ref{eq02})) allows us to determine directly this parameter but only in the situation when the inclination of the BLR is certain. Polarimetric observations have preference because the value of the polarization degree is directly associated with the inclination angle value, especially, for the standard Chandrasekhar-Sobolev theory of the generation of polarization in the plane-parallel atmosphere. Unfortunately, other methods determining inclination angle of the BLR and accretion disk are considerably uncertain. For example, it is used determination of the accretion disk inclination to the line of sight for a sample of AGNs from their bulge stellar velocity dispersion, based on suggesting a Keplerian mass/velocity dispersion between AGN and host galaxies (details in \citet{marin14}). Namely the use of the bulge stellar velocity dispersion provides the difference in the real estimate of BH masses. In many cases the BH mass is estimated suggesting that the virial coefficient $f$ = 1.

Another problem associated with the determination of the BLR size. This size cannot be directly measured from single epoch spectra of AGN, which are used for estimate of BH masses. Most popular estimated of the BLR size rely on the discovery that $R_{BLR}$ scales with a certain power of continuum luminosity of the AGN. Unfortunately this scaling dependence appears slightly different in works of various authors.

For example, for Fairall~9 and some other objects we used data for BH masses from \citet{brenneman13} and \citet{reynolds13}, that are based on the estimate of the virial coefficient in Eq.(\ref{eq01}) as $f = 1.0$. For Mrk~290 \citet{feng14} produced the estimates of its mass using the specific calibration of the coefficient of single epoch spectra in the local AGN sample. This circumstance may be the reason of the difference between our and others results.

There is yet another way for determining the SMBH mass. Under the assumption that the motion of the gas in the BLR of AGNs is dominated by the gravitational influence of the black hole, one can use the virial relation (\ref{eq01}) to obtain $M_{BH}$. The radius $R_{BLR}$ is determined usually with reverberation mapping \citep{peterson04} or is estimated with the radius-luminosity relation \citep{bentz09}.

If one adopts a value for the virial coefficient of $f = 1$, the determined $M_{BH}$ is commonly called as ''the virial product'' (VP). The results of calculations of the virial product for AGNs are presented by \citet{grier13} and \citet{ho14}. Eq.(\ref{eq04}) allows to obtain the following relation between the actual value of BH mass, virial product and $\sin{i}$:

\begin{equation}
 \sin{i} = \frac{1}{2} \left(\frac{VP}{M_{BH}}\right)^{1/2}.
 \label{eq05}
\end{equation}

\noindent Determining the value of $\sin{i}$ from the polarimetric data one can derive the actual value of $M_{BH}$. Using data on VP published in \citet{grier13,ho14}, we estimated the values of BH masses for AGNs presented in Table 2. These results are presented in Table 3. They agree with the values in Table 2 within the uncertainties.

\renewcommand{\arraystretch}{1.2}
\begin{table}
\centering
\caption{The masses of SMBH in AGNs determined via measured polarization of broad $H_\alpha$ emission and virial product data. Polarization data are from \citet{smith02}. Virial product data are from \citet{ho14}.}
\begin{tabular}{lll}
\hline
Object      & $M_{BH} / M_{\odot}$                & $M_{BH} / M_{\odot}$ \\
            & (observations)                      & (from Eq.(\ref{eq05})) \\
\hline
Akn~120     & $(3.23^{+0.44}_{-0.32})\times 10^8$ & $(2.64^{+0.50}_{-0.68})\times 10^8$ \\
Fairall~9   & $(1.34^{+0.76}_{-0.36})\times 10^8$ & $(1.26^{+2.74}_{-2.55})\times 10^8$ \\
Mrk~509     & $(1.35\pm 0.12)\times 10^8$         & $(1.26 \pm 0.0053)\times 10^8$ \\
NGC~3516    & $10^{8.06\pm 0.27}$                 & $10^{8.29^{+0.04}_{-0.03}}$ \\
NGC~3783    & $10^{7.7\pm 0.11}$                  & $10^{7.57^{+0.02}_{-0.04}}$ \\
NGC~4051    & $(1.64^{+0.67}_{-0.55})\times 10^6$ & $(1.7^{+0.6}_{-0.5})\times 10^6$ \\
NGC~4593    & $(8.25^{+3.46}_{-3.06})\times 10^6$ & $(1.15^{+0.244}_{-0.221})\times 10^7$ \\
NGC~5548    & $(7.84^{+0.53}_{-0.46})\times 10^7$ & $(1.49^{+0.101}_{-0.12})\times 10^8$ \\
NGC~6814    & $(8.7^{+1.3}_{-1.6})\times 10^6$    & $(6.4^{+0.01}_{-0.01})\times 10^6$ \\
NGC~7469    & $(3.47^{+1.59}_{-1.45})\times 10^7$ & $(4.25^{+0.19}_{-0.25})\times 10^7$ \\
PG~1211+143 & $(2.19^{+2.07}_{-0.81})\times 10^8$ & $(2.17^{+0.64}_{-0.75})\times 10^8$ \\
\hline
\end{tabular}
\end{table}
\renewcommand{\arraystretch}{1.0}

\section{Conclusions}

We presented a new method for estimating the SMBH mass. The method is based on the observations of polarization of broad lines emission, suggesting that the BLR is likely to be an optically thick flattened configuration \citep{kollatschny06,kollatschny13}. In this case the virial coefficient (Eq.(\ref{eq01})) can be derived as $f = 1 / (4 \sin^2{i})$ \citep{mclure01}. The value $\sin{i}$ can be obtained from the polarization degree of radiation of optically thick plane parallel atmosphere \citep{chandra50,sobolev63}. This result allows us to obtain an independent estimate for the central supermassive black hole mass from Eq.(\ref{eq01}). The results of our calculation of the masses of SMBH for specific AGN are presented in Table 2. Comparison of our results with the results of other authors, which used fixed values of the virial coefficients, and frequently adopted the value of $f = 1$, shows significant differences for a number of AGNs. It means that there is a clear distinction of our black hole masses from the virial product value that corresponds to the value of the virial coefficient $f = 1$ \citep{ho14}.

It is important to emphasize that the polarimetric observations allow to derive the ratio $R_{BLR} / R_g$ for the situation when the BLR resides in a flattened, disk like configuration (see Eq.(\ref{eq04})).

\section*{Acknowledgments}

This work was supported by the Basic Research Program of the Presidium of the Russian Academy of Sciences P-41 and the Basic Research Program of the Division of Physical Sciences of the Russian Academy of Sciences OFN-17.

\label{lastpage}


\begin{thebibliography}{99}
\bibitem[\protect\citeauthoryear{Afanasiev et al.}{2011}]{afanasiev11} Afanasiev V. L., Borisov N. V., Gnedin Yu. N. et al., 2011, Astron.Lett., 37, 302
\bibitem[\protect\citeauthoryear{Bentz}{2009}]{bentz09} Bentz M. C., Peterson B. M., Netzler H. et al., 2009, ApJ, 697, 160
\bibitem[\protect\citeauthoryear{Brenneman}{2013}]{brenneman13} Brenneman L., 2013, Acta Polytechnica Suppl., 53, 652
\bibitem[\protect\citeauthoryear{Chandrasekhar}{1950}]{chandra50} Chandrasekhar S., 1950, Radiative Transfer, Clarendon Press, Oxford
\bibitem[\protect\citeauthoryear{Collin et al.}{2006}]{collin06} Collin S., Kawaguchi T., Peterson B. M., Vestergaard M., 2006, Astron.Astrophys., 456, 75
\bibitem[\protect\citeauthoryear{Decarli et al.}{2008}]{decarli08} Decarli R., Labita M., Treves A., Falomo R., 2008, MNRAS, 387, 1237
\bibitem[\protect\citeauthoryear{Eracleous \& Halpern}{1994}]{eracleous94} Eracleous M., Halpern J. P., 1994, ApJSS, 90, 1
\bibitem[\protect\citeauthoryear{Eracleous, Halpern \& Livio}{1996}]{eracleous96} Eracleous M., Halpern J. P., Livio M., 1996, ApJ, 459, 89
\bibitem[\protect\citeauthoryear{Feng, Shen \& Li}{2014}]{feng14} Feng H., Shen Y., Li H., 2014, ApJ, 794, 77
\bibitem[\protect\citeauthoryear{Fine et al.}{2008}]{fine08} Fine S., Groom S. M., Hopkins P. F. et al., 2008, MNRAS, 390, 1413
\bibitem[\protect\citeauthoryear{Grier et al.}{2013}]{grier13} Grier C. J., Martini P., Watson L. C. et al., 2013, ApJ, 773, 90
\bibitem[\protect\citeauthoryear{Ho \& Kim}{2014}]{ho14} Ho L. C., Kim M., 2014, ApJ, 789, 17
\bibitem[\protect\citeauthoryear{Ho, Darling \& Greene}{2008}]{ho08} Ho L. C., Darling J., Greene J. E., 2008, ApJSS, 177, 103
\bibitem[\protect\citeauthoryear{Kollatschny \& Zetzl}{2013}]{kollatschny13} Kollatschny W., Zetzl M., 2013, Astron.Astrophys., 558, 26
\bibitem[\protect\citeauthoryear{Kollatschny, Zetzl \& Dietrich}{2006}]{kollatschny06} Kollatschny W., Zetzl M., Dietrich M., 2006, Astron.Astrophys., 454, 459
\bibitem[\protect\citeauthoryear{Labita et al.}{2006}]{labita06} Labita M., Treves A., Falomo R., Uslenghi M., 2006, MNRAS, 373, 551
\bibitem[\protect\citeauthoryear{Li, Narayan \& McClintock}{2009}]{li09} Li L.-X., Narayan R., McClintock J.E., 2009, ApJ, 691, 847
\bibitem[\protect\citeauthoryear{Marin}{2014}]{marin14} Marin F., 2014, MNRAS, 441, 551
\bibitem[\protect\citeauthoryear{McLure \& Dunlop}{2001}]{mclure01} McLure R. J., Dunlop J. S., 2001, MNRAS, 327, 199
\bibitem[\protect\citeauthoryear{McLure \& Jarvis}{2002}]{mclure02} McLure R. J., Jarvis M. J., 2002, MNRAS, 337, 109
\bibitem[\protect\citeauthoryear{Onken et al.}{2004}]{onken04} Onken C.A., Ferrarese L., Meritt D. et al., 2004, ApJ, 615, 645
\bibitem[\protect\citeauthoryear{Pancoast et al.}{2013}]{pancoast13} Pancoast A., Brewer B. J., Treu T., 2013, AAS Meeting 221, 309.08
\bibitem[\protect\citeauthoryear{Peterson et al.}{2004}]{peterson04} Peterson B. M., Ferrarese L., Gillbert K. M. et al., 2004, ApJ, 613, 682
\bibitem[\protect\citeauthoryear{Peterson \& Wandel}{1999}]{peterson99} Peterson B. M., Wandel A., 1999, ApJ, 521, L95
\bibitem[\protect\citeauthoryear{Reynolds}{2013}]{reynolds13} Reynolds C. S., 2013, arXiv:1302.3260
\bibitem[\protect\citeauthoryear{Shakura \& Sunyaev}{1973}]{shakura73} Shakura N. I., Sunyaev R. A., 1973, Astron.Astrophys., 24, 337
\bibitem[\protect\citeauthoryear{Shankar et al.}{2012}]{shankar12} Shankar F., Marull F., Mather S. et al., 2012, Astron.Astrophys., 540, A23
\bibitem[\protect\citeauthoryear{Silant'ev et al.}{2010}]{silantev10} Silant'ev N. A., Piotrovich M. Yu., Gnedin Yu. N., Natsvlishvili T. M., 2010, Astron.Rep., 54, 974
\bibitem[\protect\citeauthoryear{Smith et al.}{2002}]{smith02} Smith J. E., Young S., Robinson A. et al., 2002, MNRAS, 335, 773
\bibitem[\protect\citeauthoryear{Sobolev}{1963}]{sobolev63} Sobolev V. V., 1963, A treatise on radiative transfer, Princeton N.J., Van Nostrand
\bibitem[\protect\citeauthoryear{Tremaine \& Davis}{2014}]{tremaine14} Tremaine S., Davis S. W., 2014, MNRAS, 441, 1408
\bibitem[\protect\citeauthoryear{Vestergaard \& Peterson}{2006}]{vestergaard06} Vestergaard M., Petersen B. M., 2006, ApJ, 641, 689
\bibitem[\protect\citeauthoryear{Wang et al.}{2009}]{wang09} Wang J.-G., Dong X.-B., Wang T.-G. et al., 2009, ApJ, 707, 1339
\bibitem[\protect\citeauthoryear{Winter et al.}{2010}]{winter10} Winter L. W., Lewis K. T., Koss M. et al., 2010, ApJ, 710, 503
\bibitem[\protect\citeauthoryear{Woo et al.}{2014}]{woo14} Woo J.-H., Kim J.-G., Park D. et al., 2014, J.Korean Astron.Soc., 47, 167
\bibitem[\protect\citeauthoryear{Wu \& Han}{2001}]{wu01} Wu X. B., Han J. L., 2001, ApJ, 561, L59
\bibitem[\protect\citeauthoryear{Zhang \& Wu}{2002}]{zhang02} Zhang T.-Z., Wu X.-B., 2002, ChJAA, 2, 487
\bibitem[\protect\citeauthoryear{Ziolkowski}{2008}]{ziolkowski08} Ziolkowski J., 2008, Chinese Journal of Astronomy \& Astrophysics Suppl., 8, 273
\end{thebibliography}
\end{document}